# Overview of neutrino electromagnetic properties 2022

**Alexander Studenikin**[a,b,][*]

[a]*Department of Theoretical Physics, Lomonosov Moscow State University,
119992 Moscow, Russia*

  *E-mail:* studenik@srd.sinp.msu.ru, a-studenik@yandex.ru

[b] *National Centre for Physics and Mathematics,
  Sarov, Nizhny Novgorod Region, Russia*

Continuing the discussion of the problem of electromagnetic properties of neutrinos, in this note we present additional information on this problem and focus on selected issues that have been developed recently, after the publication of our previous brief review on this topic [1].

---

[*]Speaker

The question of the electromagnetic properties of neutrinos and, in particular, the possibility of measuring the magnetic moment, was raised [2] even before the experimental discovery of neutrinos. This is a clear confirmation of the frequently occurring statement that the electromagnetic properties of neutrinos really open a window to new physics [3].

Recall that the entirety of the electromagnetic properties of neutrinos are embodied by the effective vertex which in general contains the whole set of the neutrino electromagnetic characteristics. In the most general form the neutrino electromagnetic vertex function $\Lambda_\mu^{ij}(q)$ can be expressed [3] in terms of four form factors

$$\Lambda_\mu^{ij}(q) = \left(\gamma_\mu - q_\mu \slashed{q}/q^2\right)\left[f_Q^{ij}(q^2) + f_A^{ij}(q^2)q^2\gamma_5\right] - i\sigma_{\mu\nu}q^\nu\left[f_M^{ij}(q^2) + if_E^{ij}(q^2)\gamma_5\right], \quad (1)$$

where $\Lambda_\mu(q)$ and form factors $f_{Q,A,M,E}(q^2)$ are $3 \times 3$ matrices in the space of massive neutrinos. In the case of coupling with a real photon ($q^2 = 0$) the form factors $f(q^2)$ provide four sets of neutrino electromagnetic characteristics: 1) the electric millicharges $q_{ij} = f_Q^{ij}(0)$, 2) the dipole magnetic moments $\mu_{ij} = f_M^{ij}(0)$, 3) the dipole electric moments $\epsilon_{ij} = f_E^{ij}(0)$ and 4) the anapole moments $a_{ij} = f_A^{ij}(0)$. The expression (1) for $\Lambda_\mu^{ij}(q)$ is applicable for Dirac and Majorana neutrinos. However, a Majorana neutrino does not have diagonal electric charge and dipole magnetic and electric form factors, only a diagonal anapole form factor can be nonzero. At the same time, a Majorana neutrino can also have nonzero off-diagonal (transition) form factors.

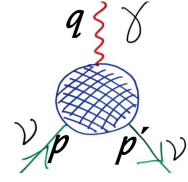

**Figure 1:** A neutrino effective one-photon coupling.

Until recently the most stringent terrestrial constraints on the effective neutrino magnetic moments have been obtained with the reactor antineutrinos: $\mu_\nu \leq 2.9 \times 10^{-11}\mu_B$ (GEMMA Collaboration [4]), and solar neutrinos: $\mu_\nu \leq 2.8 \times 10^{-11}\mu_B$ (Borexino Collaboration [5]). Both these constraints are obtained with investigations of the elastic scattering of a flavour neutrino $\nu_l$ (or an antineutrino $\bar{\nu}_l$) on an electron at rest, $\nu_l + e^- \to \nu_l + e^-$, $l = e, \mu, \tau$.

A new phase of the GEMMA project on the search for the magnetic moment of neutrinos is included into a new project, called $\nu GEN$, that is now underway at the Kalinin Power Plant aimed at the detection of coherent Neutrino–Ge Nucleus elastic scattering. It is also expected that this experiment will further increase sensitivity to the neutrino magnetic moment and will reach the level of $\mu_{\nu_e} \sim (5-9) \times 10^{-12}\mu_B$. The first release of the $\nu GEN$ experimental data has just been published [6].

It has been shown [7] that it is possible to obtain a record limit on the effective magnetic moment of neutrinos, namely $\mu_\nu^{eff} < 6.2 \times 10^{-12}\mu_B$ at 90% $CL$, using the elastic neutrino-electron scattering data published by the LUX-ZEPLIN collaboration. This limit supersedes the previous best one set by the Borexino Collaboration by almost a factor of 5 and it rejects by more than 5 the hint of a possible neutrino magnetic moment $\mu_\nu^{eff} < (1.4 - 2.9) \times 10^{-11}\mu_B$ at 90% $CL$ found by the XENON1T Collaboration [8].

Recently a combine analysis [9] of the coherent elastic neutrino-nucleus scattering (CE$\nu$NS) released by the COHERENT Collaboration and the Dresden-II reactor experiment provides a new upper limit on the electron charge radius and also significantly improves the other CE$\nu$NS-related limits on the neutrino electric charge and magnetic moment.

In [10] the $\nu - e$ scattering at the DUNE near detector is used to study the electromagnetic properties of neutrinos and the constraints that can be obtained on the magnetic moment, the electric millicharge and the charge radius of neutrinos are discussed. The performed analysis shows that DUNE will be not able to reach GEMMA's sensitivity to the magnetic moment of $\nu_e$, as well as the existing constraints on $q_{\nu_e}$ from reactors which is at the $O(10^{-12} e_0)$ level [11, 12]. However DUNE will be able to extend sensitivity to $q_{\nu_\mu}$ by improving over COHERENT's existing constraints by around two orders of magnitude. It will be also possible to consider inelastic neutrino-electron scattering at DUNE and derive constraints on active-sterile transition magnetic moments which have recently attracted significant attention (see, for instance, [13, 14]).

Neutrino transition magnetic moments can influence the neutrino flavour evolution inside a galactic SN. By simulating the event spectra for DUNE and Hyper-Kamiokande experiments it has been found [15] that under certain conditions the transition magnetic moments can be probed by these experiments on two or three orders of magnitude better than the current terrestrial and astrophysical bounds (see, for instance, [1]).

In [16, 17] we have proposed an experimental setup to observe coherent elastic neutrino-atom scattering using electron antineutrinos from tritium decay and a liquid helium target. The estimated sensitivity of this apparatus to the electron neutrino magnetic moment is about $\mu_\nu \leq 7 \times 10^{-13} \mu_B$, that is approximately two orders of magnitude smaller than the current experimental limits from GEMMA and Borexino. A corresponding experiment involving the use of an intense $1 kg$ tritium antineutrino source is currently being prepared in the framework of the research program of the National Center for Physics and Mathematics in Sarov. The status of the project implementation is given in [18].

Just very recently a novel way to probe the neutrino magnetic and electric moments by using the atomic radiative emission of neutrino pair has been discussed in [19]. It is claimed that the sensitivity for the neutrino magnetic moment at the level $(1.5 \sim 3.5) \times 10^{-11} \mu_B$ is reachable.

The work is supported by the Russian Science Foundation under grant No.22-22-00384.

## References

[1] A. Studenikin, *Overview of neutrino electromagnetic properties*, *PoS* **CORFU2021** (2022) 057.

[2] C. L. Cowan, F. Reines and F. B. Harrison, *Upper limit on the neutrino magnetic moment*, *Phys. Rev.* **96** (1954) 1294.

[3] C. Giunti and A. Studenikin, *Neutrino electromagnetic interactions: A window to new physics*, *Rev. Mod. Phys.* **87** (2015) 531.

[4] A. Beda, V. Brudanin, V. Egorov D. V. Medvedev, V. S. Pogosov, M. V. Shirchenko et al., *The results of search for the neutrino magnetic moment in GEMMA experiment*, *Adv. High Energy Phys.* **2012** (2012) 350150.

[5] M. Agostini *et al.* [Borexino Collaboration], *Limiting neutrino magnetic moments with Borexino Phase-II solar neutrino data*, *Phys. Rev. D* **96** (2017) 091103.

[6] I. Alekseev et al. [νGeN], *First results of the νGeN experiment on coherent elastic neutrino-nucleus scattering*, Phys. Rev. D **106** (2022) L051101.

[7] M. Atzori Corona, W. M. Bonivento, M. Cadeddu, N. Cargioli and F. Dordei, *New constraint on neutrino magnetic moment from LZ dark matter search results*, [arXiv:2207.05036 [hep-ph]].

[8] E. Aprile et al., [XENON Collaboration], *Excess electronic recoil events in XENON1T*, Phys. Rev. D **102** (2020) 072004.

[9] M. Atzori Corona, M. Cadeddu, N. Cargioli, F. Dordei, C. Giunti, Y. F. Li, C. A. Ternes and Y. Y. Zhang, *Impact of the Dresden-II and COHERENT neutrino scattering data on neutrino electromagnetic properties and electroweak physics*, JHEP **09** (2022) 164.

[10] V. Mathur, I. M. Shoemaker and Z. Tabrizi, *Using DUNE to shed light on the electromagnetic properties of neutrinos*, JHEP **10** (2022) 041.

[11] A. Studenikin, *New bounds on neutrino electric millicharge from limits on neutrino magnetic moment*, Europhys.Lett. **107** (2014) 21001.

[12] A. Parada, *Constraints on neutrino electric millicharge from experiments of elastic neutrino-electron interaction and future experimental proposals involving coherent elastic neutrino-nucleus scattering*, Adv. High Energy Phys. **2020** (2020) 5908904.

[13] T. Schwetz, A. Zhou and J. Y. Zhu, *Constraining active-sterile neutrino transition magnetic moments at DUNE near and far detectors*, JHEP **21** (2020) 200.

[14] P. D. Bolton, F. F. Deppisch, K. Fridell, J. Harz, C. Hati and S. Kulkarni, *Probing active-sterile neutrino transition magnetic moments with photon emission from CEvNS*, Phys. Rev. D **106** (2022) 035036.

[15] S. Jana, Y. P. Porto-Silva and M. Sen, *Exploiting a future galactic supernova to probe neutrino magnetic moments*, JCAP **09** (2022) 079.

[16] M. Cadeddu, F. Dordei, C. Giunti, K. Kouzakov, E. Picciau, A. Studenikin, *Potentialities of a low-energy detector based on $^4$He evaporation to observe atomic effects in coherent neutrino scattering and physics perspectives*, Phys. Rev. D **100** (2019) 073014.

[17] E. Picciau, M. Cadeddu, F. Dordei, C. Giunti, K. A. Kouzakov and A. I. Studenikin, *New process in superfluid $^4$He detectors: The coherent elastic neutrino-atom scattering*, PoS **ICHEP2020** (2021) 211.

[18] M. Cadeddu, G. Donchenko, F. Dordei, C. Giunti, K. Kouzakov, B. Lubsandorzhiev, A. Studenikin, V. Trofimov, M. Vyalkov, A. Yukhimchuk, *A proposal for experiment with high-intensity tritium neutrino source in Sarov: The search for coherent elastic neutrino-atom scattering and neutrino magnetic moment*, PoS **ICHEP2022** (2022) 591.

[19] S. F. Ge and P. Pasquini, *Unique probe of neutrino electromagnetic moments with radiative pair emission,* [arXiv:2206.11717 [hep-ph]].